\documentstyle[12pt,epsfig]{article}
\def\mb#1{\mbox{\boldmath $#1$}}
\def\beq{\begin{equation}}
\def\eeq{\end{equation}}
\def\qv{{\mb q}}

\def\qt{{\mb q}_\perp}
\def\qbart{\bar {\mb q}_\perp}
\def\pv{{\mb p}}

\def\rv{{\mb r}}
\def\pt{{\mb p}_\perp}
\def\xu{\hat{\mb x}}
\def\yu{\hat{\mb y}}
\def\zu{\hat{\mb z}}
\def\pol{{\mb P}}
\def\polt{{\mb P}_\perp}
\def\Et{E_\perp}
\def\Lv{{\mb L}}

\def\sv{{\mb s}}

\def\Jv{{\mb J}}
\def\3p0{$^3\!P_0$}
\def\EF{{\cal E}}
\def\D{\partial}

\def\PRep{Phys.\ Rep.\ }
\def\PR{Phys.\ Rev.\ }
\def\PRL{Phys.\ Rev.\ Lett.\ }
\def\ZP{Z.\ Phys.\ }
\def\NP{Nucl.\ Phys.\ }
\def\PL{Phys.\ Lett.\ }
\begin{document}
\begin{titlepage}
\pagestyle{empty}
\vspace*{4cm}
\begin{center}
{\large\bf TRANSVERSE POLARIZATION OF QUARK PAIRS \\
            CREATED IN STRING FRAGMENTATION }
\vspace{1.1cm}\\
{\sc X.~Artru}{$^a$},
{\sc J.~Czy\.zewski}{$^b$}
\vspace{0.3cm}\\
$^a${\it Institut de Physique Nucl\'eaire de Lyon, IN2P3-CNRS
et Universit\'e Claude Bernard, France}
\vspace{0.3cm}\\
$^b${\it Institute of Physics, Jagellonian University, 
Cracow, Poland}
\end{center}
\vspace{0.0cm}

\begin{abstract}  
Classical arguments predict that the quark and the antiquark of a pair created 
during string fragmentation are both transversely polarized in the direction 
of $\zu\times\qt$, where $\zu$ is the direction of the pull exerted by the 
string on the antiquark and $\qt$ ($-\qt$) is the transverse momentum of the 
quark (antiquark).  The existence of this effect at the quantum-mechanical 
level is investigated by considering two analogous processes involving the 
tunnel effect in a strong field~: (1) dissociation of the positronium atom 
(2) electron pair creation. In case (1) the positronium is taken in the 
\3p0 state to simulate the vacuum quantum numbers $J^{PC}=0^{++}$.  Using 
the nonrelativistic WKB method, the final electron and positron are indeed 
found to be transversely polarized along $\zu\times\pt$.  On the contrary, 
case (2), treated with the Dirac equation, shows no correlation between 
transverse polarization and transverse momentum both when the field is 
uniform and when it depends on $z$ and $t$. The pair is nevertheless produced 
in a triplet spin state.  The difference between these two results and their 
relevance to transverse spin asymmetry in inclusive reactions is discussed.
\end{abstract} 
\vspace{0.5cm}
\noindent
{\sf TPJU 9/98}\\
{\sf LYCEN-9834}\\
{\sf May 1998}
\end{titlepage}
\section{Introduction}

The experimental observation of single-spin asymmetries in inclusive hadron 
production at high energy \cite{SSA} have been tentatively explained by 
various models~: Thomas precession \cite{ThP}, P-wave orbitals \cite{PWO}, 
Regge exchange \cite{Soffer,Barni}, semi-classical string mechanism 
\cite{Andersson,ACY}.  The asymmetric part of the cross-section is of the 
form $A \ \polt\cdot(\zu\times\hat\pt)$ where $\zu$ is the collision axis, 
$\pt$ the transverse momentum of the produced particle, $\polt$ its 
transverse polarization ({\it transversity}) or that of one of the colliding 
baryons.  The "hat" denotes a unitary vector: $\hat\pt\equiv \pt/|\pt|$. A 
similar effect correlating the transversity of the leading quark of a jet 
with the transverse momentum of one of the fastest particles was predicted 
by Collins \cite{Collins}.  This effect, if confirmed experimentally, would 
serve as a transverse ``quark polarimeter''. It was used in Ref.~\cite{ACY} 
to explain single spin asymmetry in inclusive meson production.

In this paper we start from a popular string hadronization picture 
\cite{Casher,Gurvich,Glendenning}, in which quark-antiquark pairs are 
produced from the string by a tunneling mechanism analogous to the Schwinger 
mechanism for pair creation in a strong homogeneous electric field [12--20].
% \cite{Schw,Brezin,Wang,Martin,Herrmann,Schonfeld,Pavel,Kluger,Smolyanskii}.
%
%%%%%%%%%%%%%%%%%%%%%%%%%%%%%%%%%%%%%%%%%%%%%%%%%%%%%%%%%%
\begin{figure}[ht]
\begin{center}
{\epsfig{file=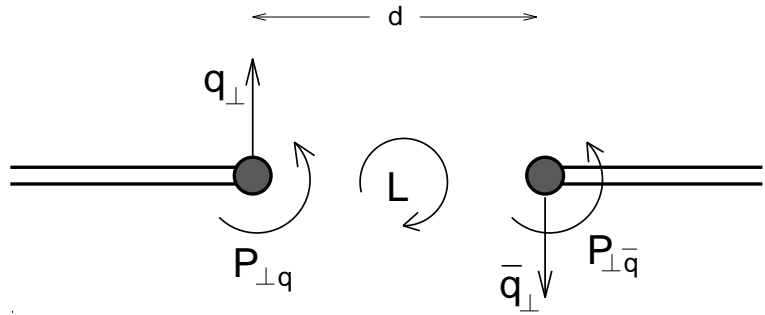,width=10cm}}
\label{fig1}
\parbox{13cm}{
\caption{\footnotesize
Semi-classical string mechanism of quark polarization. The orbital
angular momentum of the $\bar qq$ pair is compensated by the spin
of $q$ and $\bar q$, thereby causing the correlation between spin and 
transverse momentum of the quark and the antiquark.}
}
\end{center}
\end{figure}
%%%%%%%%%%%%%%%%%%%%%%%%%%%%%%%%%%%%%%%%%%%%%%%%%%%%%%%%%%
%
This picture accounts rather well for the exponential cut-off in $\pt$, the 
relative suppression of strange quarks and the almost complete suppression 
of heavy quarks.  Let us recall the semi-classical arguments \cite{Andersson} 
for a transverse  polarization of the quark and the antiquark (see Fig.~1)~:

\begin{itemize}
\item[-] 
the quark and the antiquark come from a pair fluctuation like those which
occur in ordinary vacuum.  At zero separation, the pair has zero total 
energy-momentum.  In particular, quark and antiquark have opposite transverse 
momentum $\qt$ and $\qbart$.  In the vacuum case, the pair stays virtual and 
disappears after a time of the order of the quark Compton wave length.  In 
the string case, the linear mass density $\kappa\simeq$ 1~GeV/fm of the 
string is converted into energy of the pair, which becomes real at a 
longitudinal separation $d=2\Et/\kappa$, where $\Et = (m^2+\qt^2)^{1/2}$ 
is the quark (or antiquark) transverse energy.
\item[-] The orbital angular momentum 
\beq
\Lv = d \ \zu\times\qbart
\eeq
is compensated, at least partly, by the spins of the quark and the antiquark.
Assuming equal polarization
$\pol \equiv 2 \, <\sv_q> \, = 2 \, <\sv_{\bar q}> $ for the quark and the
antiquark, we have therefore
\beq
\pol = - \Lv \ f(L),
\eeq
where $f(L)$ is a reduction factor insuring that $|\pol|$ is smaller 
than unity, {\it e.g.}, $f(L) = (1+L)^{-1}$. 
\end{itemize}

\noindent
To summarize, the polarizations of the quark and of the antiquark 
are of the form
\beq
\pol^q = \pol^{\bar q} = - A(q_\perp) \ \zu\times\hat\qv_\perp
\,,
\label{AP}
\eeq
where
\beq
A(q_\perp)= L \, f(L) \ \le {\rm min}\{L,1\}
\label{AP'}
\eeq
is the {\it analysing power} of the mechanism and
\beq
\qquad L=2\kappa^{-1} \, q_\perp \, (m^2+q_\perp^2)^{1/2}
\,.
\eeq
The compensation between $\Lv$ and $\sv_q+\sv_{\bar q}$ is further motivated
by the phenomenological success of the ``\3p0'' model of quark pair creation 
in hadronic decay \cite{3P0}. This model assumes that the $(q \,\bar q)$ 
pair is created in the \3p0 spin state, therefore having the vacuum quantum 
numbers $J^{PC} = 0^{++}$ (by contrast, a model in which the pair comes from 
one gluon gives $J^{PC} = 1^{--}$).

The string, as well as the constant electric field, is not rotationally 
invariant and therefore the total angular momentum $\Jv=\Lv+\sv_q+\sv_{\bar q}$ 
is not conserved during tunneling. This is a weak point of the classical 
model reviewed above.  C- and P- quantum numbers are also not separately 
conserved. Therefore it is desirable to check the existence of the transverse
polarization effect in a true quantum-mechanical model.  Although too 
difficult at present in the string theory, this is quite possible for the 
problem of pair creation in strong homogeneous field. To begin with (in 
section 2) we will consider the nonrelativistic process of the dissociation 
of a positronium atom, which we assume to be in the \3p0 state. The 
relativistic case of pair creation in a field $\vec\EF(t,z)$ parallel to 
the $\zu$ axis (independent of $x$ and $y$ but not necessarily on $t$ and $z$)
will be considered in section 3, using the Dirac hole theory for the positron.
The different results obtained in these two cases will be discussed in 
Section 4.

\section{Positronium dissociation}

The nonrelativistic $e^+\,e^-$ system in a constant electric field 
is governed by the Hamiltonian
$$
H={\pv_+^2\over2m} + {\pv_-^2\over2m} - {\alpha\over|\rv_+-\rv_-|} 
+ e\, \vec\EF\cdot(\rv_--\rv_+)
$$
\beq
= {{\bf p}_{tot}^2\over4m} + {\pv^2\over m} - {\alpha\over r} - Fz
\equiv K_{barycentre} + K_r + V_c - Fz
\,.
\label{hamil}
\eeq
The motion of the barycentre can be separated from the relative motion and 
from now on we will consider only the latter, governed by the three last
terms of Eq.~(\ref{hamil}).
$\rv=\rv_+-\rv_-$ and $\pv=(\pv_+-\pv_-)/2$ 
are the relative position and momentum.  We take $F \equiv e \EF_z >0$.  The 
potential $V_c-Fz$ is shown in Fig.~2.  Note that the Hamiltonian is 
spin-independent, therefore cannot produce spin effects by itself.  However, 
we will assume that the initial state of the pair is a \3p0 positronium 
(corresponding to the vacuum quantum numbers)~:
%
%%%%%%%%%%%%%%%%%%%%%%%%%%%%%%%%%%%%%%%%%%%%%%%%%%%%%%%%%%
\begin{figure}[ht]
\begin{center}
\mbox{\epsfig{file=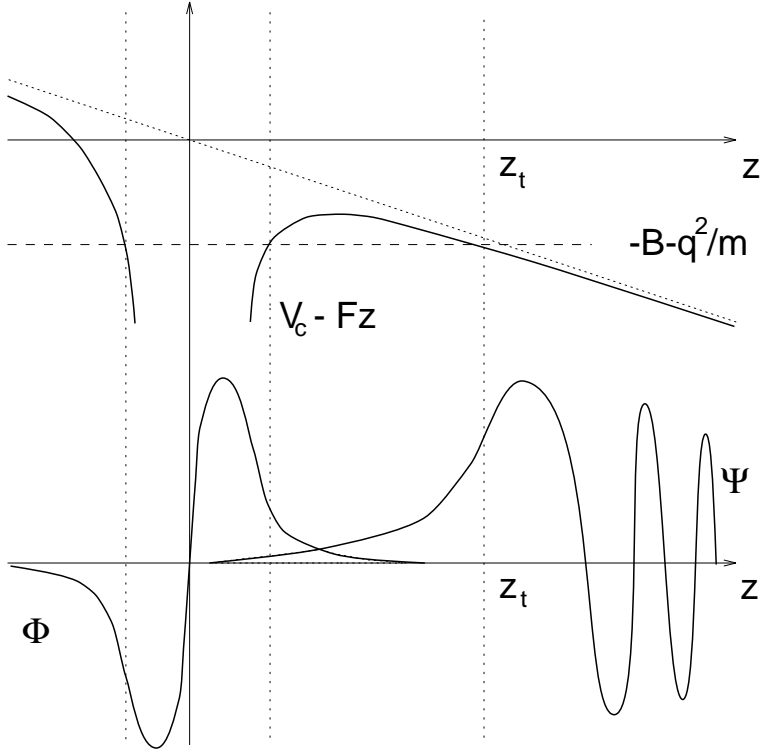,width=10cm}}
\label{fig2}
\parbox{13cm}{
\caption{\footnotesize
Positronium dissociation in constant electric field. The top curve is the 
superposition of the Coulomb potential and the external electric potential. 
The classically allowed region for a given transverse momentum $q$ is limited 
by the horizontal dashed line.  The bottom curves are the $p$ wave function 
of the positronium, $\Phi$, and the wave function of the free solution, 
$\Psi$ (restricted to the $z$ axis).  The overlap of $\Phi$ and $\Psi$ is 
responsible for the tunneling.}
}
\end{center}
\end{figure}
%%%%%%%%%%%%%%%%%%%%%%%%%%%%%%%%%%%%%%%%%%%%%%%%%%%%%%%%%%
%
\beq 
\Phi(\rv) = f(r) 
\ \left[ 
\ Y_1^{-1}(\hat\rv) \ |+1\rangle
\ - Y_1^0(\hat\rv) \ |0\rangle
\ + Y_1^{+1}(\hat\rv) \ |-1\rangle
\ \right]
\,,
\label{Phi}
\eeq
where the kets denote the three different triplet spin states,
\beq
|+1\rangle = |\uparrow\uparrow\rangle\,, \qquad
|0\rangle = (|\uparrow\downarrow\rangle + |\downarrow\uparrow\rangle)/\sqrt{2}
\,,\qquad |-1\rangle = |\downarrow\downarrow\rangle
\,.
\eeq
In this way, the orbital motion and the spin are entangled.  This state is a 
bound eigenstate of $K_r+V_c$ with energy $-B$.  After turning on the external 
electric field, the relative wave function will eventually migrate toward 
$z=+\infty$ by tunnel effect, which means that the positron runs toward 
$z=+\infty$ and the electron runs toward $z=-\infty$. The pair remains in the 
spin-triplet subspace. We chose $\xu$ as the spin quantization axis and are 
interested in the relative probabilities to obtain the different final spin 
states $|S_x\rangle$, with $S_x \equiv s^+_x + s^-_x$, for a given final 
transverse momentum $\qt = q \, \yu$. The corresponding asymptotic state is 
an eigenstate of $K_r-Fz$, again with energy $-B$, and its wave function is
\beq
\Psi(\rv) = \psi(\rv) \ |S_x\rangle = e^{iqy} \ g(z-z_t) \ |S_x\rangle
\eeq
where $g$ is the Airy function, solution of the one-dimensional Schr\"odinger 
equation
\beq
\left[ {p_z^2 / m} - F\,(z-z_t) \right] \, g(z) = 0
\eeq
and $z_t=(B + q^2/m)/F$ is the classical turning point (see Fig.~2). 
Here we give a heuristic proof and an estimation\begin{footnote}
%------------------------------------------------------
{A more rigourous treatment could be done using 
the method of Landau \& Lifshitz 
({\it Quantum Mechanics} \cite{Landau}),
for hydrogen dissociation in a strong electric field.}
\end{footnote}
%-----------------------------------------------------
of the spin asymmetry~:
\begin{itemize}
\item[-] We assume that the tunneling length $z_t\sim m\alpha^2/(16 F)$ 
is much larger than the radius $\sim 8/(m\alpha)$ of the bound state.
Near the bound state, we can use the WKB approximation for $g$~:
\beq
g(z-z_t) \sim e^{\lambda (z-z_t)}
\eeq
with $\lambda = (mB+q^2)^{1/2}$.
\item[-] Near the origin, $\psi(\rv)$ can be expanded in partial waves~:
\beq
\psi(\rv) \simeq e^{i\pv\cdot\rv}
= 4\pi\sum_{l,m} i^l \ j_l(pr) \ (-1)^m \ Y_l^{-m}(\hat\pv) \ Y_l^m(\hat\rv)
\eeq
with $\pv = q\yu -i\lambda\zu$, $\,p=i(\lambda^2-q^2)^{1/2}$,
$\,\hat\pv=\pv/p$.
We assume that tunneling couples mainly the components of $\Phi$ and $\Psi$
with the same $Y^m_l(\hat\rv)$, and the tunneling amplitude is proportional to 
the coefficient of this harmonic (this is intuitive if we consider the inverse 
process of trapping an initially free particle into the Coulomb potential 
well). The $l=1$ terms of $\psi$ are proportional to 
$$
j_1(pr) \ |\pv^2|^{-1/2}
\ [\ -(p_y-ip_z) \, Y^{+1}_1(\hat\rv)  + (p_y+ip_z) \, Y^{-1}_1(\hat\rv) \ ]
$$
\beq
= j_1(pr) \ (\lambda^2-q^2)^{-1/2}
\ [\ (\lambda-q) \, Y^{+1}_1(\hat\rv) + (\lambda+q) \, Y^{-1}_1(\hat\rv) \ ]
\,,
\eeq
\end{itemize}
Comparing with Eq.~(\ref{Phi}) we find that the tunneling amplitudes squared 
are in the ratio 
\beq
|T(S_x = +1)|^2 \ : \ |T(S_x = 0)|^2 \ : \ |T(S_x = -1)|^2 
\ = \ |\lambda+q|^2 \ : \ 0 \ : \ |\lambda-q|^2
\eeq
Note the vanishing of $T(S_x = 0)$. It happens because the second term of 
$\Phi$ is odd in $x$ and cannot tunnel to $\Psi$, which is even in $x$ 
(orbital $x$-parity is a symmetry of the problem). The polarization of the 
electron and the positron are equal and given by 
\beq
\pol = {|T(+1)|^2 - |T(-1)|^2 \over |T(+1)|^2 + |T(-1)|^2} \ \xu
= -\,\,{2 \sqrt{mB+q^2} \over mB+2q^2 } \ \zu \times \qt
\eeq
We see that the polarization of the created particle is of the form 
(\ref{AP}), (\ref{AP'}) and has the same sign as predicted by the classical 
string arguments. Classical trajectories leading to the positronium 
dissociation shown in Fig.~3 explain this fact intuitively.
%
%%%%%%%%%%%%%%%%%%%%%%%%%%%%%%%%%%%%%%%%%%%%%%%%%%%%%%%%%%%%%%%%
\begin{figure}[ht]
\begin{center}
\mbox{\epsfig{file=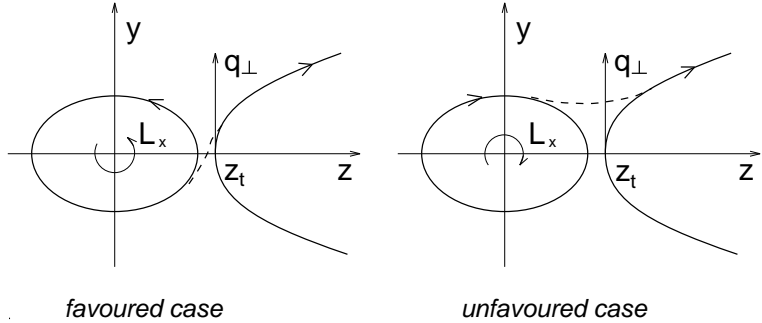,width=10cm}}
\label{fig3}
\parbox{13cm}{
\caption{\footnotesize
Classical trajectories of the positronium dissociation for the two cases 
$L_x=+1$ and $L_x=-1$.  The dashed lines are (classically forbidden) 
tunneling trajectories.}
}
\end{center}
\end{figure}
%%%%%%%%%%%%%%%%%%%%%%%%%%%%%%%%%%%%%%%%%%%%%%%%%%%%%%%%%%%%%%%%

\section{Pair creation in strong field}

In a static constant electric field, electron-positron pairs are created 
spontaneously in vacuum, the positron running in the field direction and 
the electron in the opposite direction.  In the Dirac hole theory, this 
process is interpreted as tunneling of the electron from the Dirac sea in 
one half-space to the upper continuum in the opposite half-space, without 
changing its total energy, as shown in Fig.~4 \cite{Schw}. 
%
%%%%%%%%%%%%%%%%%%%%%%%%%%%%%%%%%%%%%%%%%%%%%%%%%%%%%%%%%%%%%%%
\begin{figure}[ht]
\begin{center}
\mbox{\epsfig{file=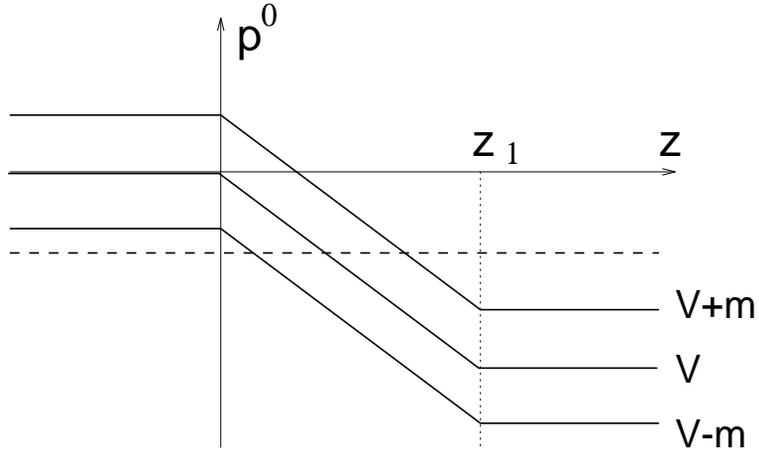,width=10cm}}
\label{fig4}
\parbox{13cm}{
\caption{\footnotesize
The Dirac sea distorted by constant electric field between 0 and $z_1$.  A 
negative-energy electron on the left-hand side can reach the upper continuum 
of the right-hand side by tunneling trough the forbidden band, becoming 
physical electron. The hole created on the left-hand side is the physical 
positron. The dashed line represent the energy of the tunneling wave function.}
}
\end{center}
\end{figure}
%%%%%%%%%%%%%%%%%%%%%%%%%%%%%%%%%%%%%%%%%%%%%%%%%%%%%%%%%%%%%%%
%
We will study the more general case of a time- and z-dependent field
$\vec\EF(t,z)$ parallel to $\zu$ and consider Dirac wave functions of definite
transverse momentum parallel to the $\yu$ axis~: $\pt\equiv(p_x,p_y)=(0,q)$.
Discarding the trivial $y$-dependence in $\exp(iqy)$, the Dirac equation reads
\beq
[\,i\D_t+eA_0 - \alpha_z(i\D_z+eA_z) - q\alpha_y -m\beta\,]\ \psi(t,z)=0
\,.
\eeq
Calculations will be simpler using the light-cone coordinates
\beq
\eta=(t+z)/2 \,, \qquad \xi=(t-z)/2
\eeq 
\beq
\D_\eta=\D_t+\D_z \,, \qquad \D_\xi=\D_t-\D_z
\eeq
\beq
A_\eta=A_0+A_3 =A^0-A^3 \,, \qquad A_\xi=A_0-A_3 =A^0+A^3 
\eeq
Furthermore we choose the spinorial representations of the Dirac matrices
\beq
\alpha^i=\pmatrix{
\sigma^i & 0 \cr
0 & -\sigma^i \cr }
\,,\quad
\beta=\pmatrix{
0 & I \cr
I & 0 \cr }
\,,\quad
\Sigma^i=\pmatrix{
\sigma^i & 0 \cr
0 & \sigma^i \cr }
\,,
\eeq
where $\sigma^i$ are the Pauli matrices. The Dirac equation becomes 
$$
[\,\alpha^\eta \, (i\D_\eta+eA_\eta)  + \alpha^\xi \, (i\D_\xi+eA_\xi) 
- q\alpha_y -m\beta\,]\ \psi(\eta,\xi) \equiv
$$
\beq
\pmatrix{
i\D_\eta+eA_\eta & iq & -m & 0 \cr
-iq & i\D_\xi+eA_\xi & 0 & -m  \cr
-m & 0 & i\D_\xi+eA_\xi & -iq  \cr
0 & -m & iq & i\D_\eta+eA_\eta \cr }
\psi(\eta,\xi) = 0
\eeq
This equation is invariant under the transformation 
$\psi_1 \leftrightarrow \psi_4$, 
$\,\psi_2 \leftrightarrow \psi_3$, 
which is performed by the matrix $\beta\,\Sigma_x=\gamma^x\gamma^5$. This 
matrix commutes with the Hamiltonian and has eigenvalues $\pm1$. For a particle 
at rest, $\beta\,\Sigma_x = 2 s_x$.  For a particle with nonzero $p_y$ and 
$p_z$, it is the $x$-component of the transversity operator. For the states 
under consideration, the $\beta\,\Sigma_x$ transformation is equivalent to 
the parity about the (y,z) plane, 
\beq
P_{yz} = e^{-i\pi J_x} \, P
= -i \, \Sigma_x \times \beta \times P_{\rm intrinsic} 
\times P_{yz}^{\rm orbital}  
\,,
\eeq
Since $p_x=0$, 
$\,P_{yz}^{\rm orbital}=1$ 
and $\beta\,\Sigma_x$ is equivalent to $P_{yz}$ (up to a phase factor, 
depending on the choice of the intrinsic parity).  The $\beta\,\Sigma_x$ 
invariance comes therefore from the symmetry of the problem about the (y,z) 
plane. 

Taking the transversity eigenstates
\beq
\psi^\uparrow = {1\over\sqrt{2}} \pmatrix{ F\cr G\cr G\cr F\cr }
\,,\qquad
\psi^\downarrow = {1\over\sqrt{2}} \pmatrix{ F\cr G\cr -G\cr -F\cr }
\,,
\eeq
we come to coupled differential equations
\beq
(i\D_\eta+eA_\eta) \ F = (\pm m-iq) \ G
\label{FG}
\eeq
\beq
(i\D_\xi+eA_\xi) \ G = (\pm m+iq) \ F
\label{GF}
\eeq
where $\pm$ is the sign of the transversity. 
From these we get the second order differential equations
\beq
[\,(i\D_\xi+eA_\xi)\,(i\D_\eta+eA_\eta)-M^2\,] \ F = 0
\label{F}
\eeq
\beq
[\,(i\D_\eta+eA_\eta)\,(i\D_\xi+eA_\xi)-M^2\,] \ G = 0
\,,
\label{G}
\eeq
where $M=(m^2+q^2)^{1/2}$ is the ``transverse energy'' of the electron. 
Note that Eqs.~(\ref{F}) and (\ref{G}) depend only on the transverse energy, 
not on the transversity. We can infer that there is no correlation between 
transverse spin and transverse momentum, contrarily to the case of positronium 
dissociation.  Let us check this result more carefully. Denoting by $F_{m,q}$ 
and $G_{m,q}$ the solution of Eqs.~(\ref{FG},\ref{GF}) and setting 
$m+iq=M \, e^{i\alpha}$, we have 
\beq
F_{m,q} = F_{M,0} \ e^{\mp i\alpha/2}
\,,\qquad
G_{m,q} = G_{M,0} \ e^{\pm i\alpha/2}
\,.
\label{m-q}
\eeq
These equations tell how the solution transforms under a ``rotation'' in 
the (m,q) plane (which leaves M invariant). The current 4-vector 
$(J^0,J^i)=(\psi^\dagger\psi,\psi^\dagger\alpha^i\psi)$
is given by 
\beq
J^0 = |F|^2 + |G|^2
\,,\qquad
J^x = 0
\,,\qquad
J^z = |F|^2 - |G|^2
\eeq
\beq
J^y =\  2\ {\rm Im} \, (F^*G) =\  
{2q\over M^2} \ Re\, [\, F^* \,(i\D_\eta+eA_\eta) \,F \,]
\pm {m\over M^2} \ \D_\eta |F|^2
\,.
\eeq
The last form of $J^y$ has been obtained using Eq.~(\ref{FG}). The first part 
is the ``convection'' term, the second part is the ``magnetization'' term.  
$J^0$, $\,J^z$ and the convection part of $J^y$ do not depend on the 
transversity. This confirms the absence of transverse spin effect in the
Schwinger mechanism of pair creation. The magnetization term of $J^y$ depends 
on transversity but is located at the edges of the wave packet (we can replace 
$\D_\eta |F|^2$ by $(\D_0\,J^z+\D_z\,J^0)/2$, using  current conservation) and 
is not observable by a macroscopic $e^\pm$ detector. 

To fix the idea, let us consider a homogeneous field $\vec\EF=\EF\,\zu$ 
confined in the region $0<\xi<\xi_1$.  This field corresponds to a capacitor 
moving with light velocity. We use the null-plane gauge 
\beq
A_\xi=0\,,\quad A_\eta = \cases{ 
0 & if $\xi<0$ \cr
2\,\xi\,\EF & if $0<\xi<\xi_1$ \cr 
2\,\xi_1\,\EF & if $\xi_1<\xi$ \cr 
}
\label{champ}
\eeq
At fixed light-cone momentum $p_\eta \equiv p_0+p_3 \equiv p^0-p^z$, 
we have the following solutions~:
\beq
G=e^{-ip_\eta \eta-iM^2\,\xi/p_\eta}
\,,\quad 
F= {\pm m-iq\over p_\eta} \ G 
\,,\qquad(\xi<0)
\eeq
\beq
G=e^{-ip_\eta \eta} 
\ \left( {2 \, \kappa \xi \over p_\eta} + 1 \right)^{-i{M^2\over2\kappa}}
\,,\quad 
F= {\pm m-iq \over p_\eta + 2 \, \kappa \xi} \, G 
\,,\qquad(0<\xi<\xi_1)
\label{jet}
\eeq
\beq
G=\left( {P'_\eta \over P_\eta} + 1 \right)^{-i{M^2\over2\kappa}}
\ \exp\left[-ip_\eta \eta -iM^2 \, 
{\xi-\xi_1 \over p_\eta + 2 \, \kappa \xi_1} 
\right]
\,,\quad 
F= {\pm m-iq \over p_\eta + 2 \, \kappa \xi_1} \, G 
\,,\qquad(\xi_1<\xi)
\,.
\eeq
$\kappa=e\EF$ is the electric force and $P_\eta=p_\eta$ and 
$P'_\eta= p_\eta + 2 \, \kappa \xi_1 $ are the initial and final ``mechanical'' 
(gauge invariant) light-cone momenta ($P_\mu \equiv p_\mu + e A_\mu$). 
Electrons from the Dirac sea at $\xi<0$ having light-cone momentum $P_\eta$ 
in the range $[-2\kappa \xi_1 ,0]$ become physical electrons ($P'_\eta>0$) at  
\beq
\xi_c \equiv -P_\eta/(2\kappa)
\eeq
The electron flux going through the hyperplane $\xi=$ constant being 
proportional to $J^\xi=(J^0-J^z)/2=|G|^2$, the tunneling probability is
\beq
J^\xi(\eta,\xi>\xi_c) \, / \, J^\xi(\eta,\xi<\xi_c) = e^{-\pi\,M^2/\kappa}
\,.
\label{expo}
\eeq
This result is clearly independent on spin.

Some remarks have to be made concerning the above calculation:
\begin{itemize}
\item[-] 
The last result is obtained giving a small positive imaginary part to $P_\eta$.
This corresponds to the physical condition that the field does not interact 
with the wave at $t=-\infty$.  \item[-] In such a field, the created electron 
escape the field region (at $\xi=\xi_1$) but not the positron. This can be 
seen from their classical trajectories in the field region shown in Fig.~5~:
\beq
\eta=\eta_0 \pm M/(2\kappa) \ e^r
\,,\quad
\xi= \xi_c \mp M/(2\kappa) \ e^{-r}
\,,\quad
y=y_0 + q\,r/\kappa 
\label{hyperbola}
\eeq
where $\eta_0$ and $y_0$ are free parameters and 
$r=\pm\,(\kappa/M)\times$(proper time) is the rapidity of the $e^\pm$.
%
%%%%%%%%%%%%%%%%%%%%%%%%%%%%%%%%%%%%%%%%%%%%%%%%%%%%%%%%%%%%%%%%
\begin{figure}[ht]
\begin{center}
\mbox{\epsfig{file=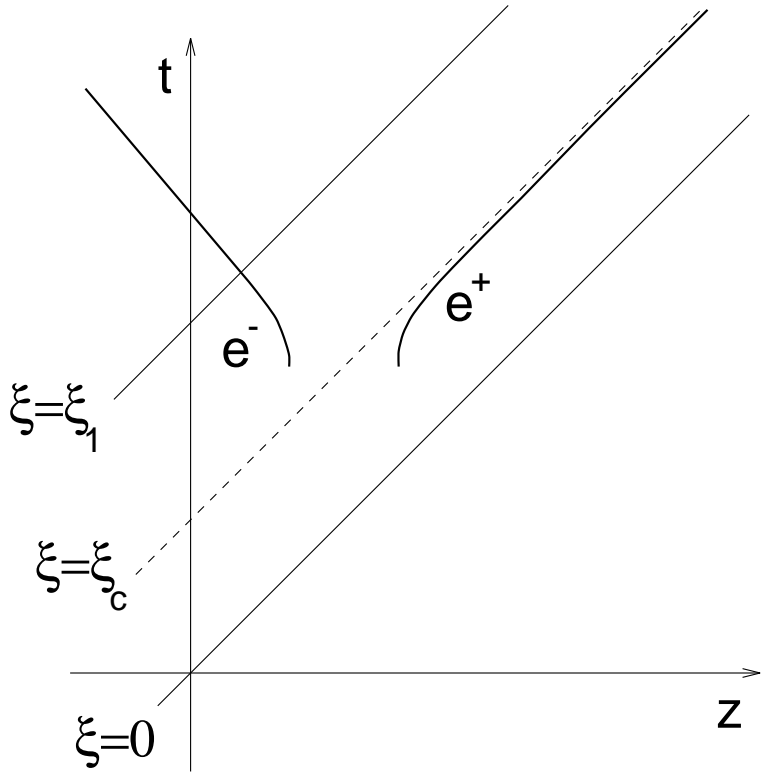,width=10cm}}
\label{fig5}
\parbox{13cm}{
\caption{\footnotesize
Classical trajectories of the electron and the positron described by Eq.~(37) 
created in the field Eq.~(31) confined in the region bounded by the two solid 
diagonal lines. The electron escapes from the field region while the positron 
remains in the field forever.}
}
\end{center}
\end{figure}
%%%%%%%%%%%%%%%%%%%%%%%%%%%%%%%%%%%%%%%%%%%%%%%%%%%%%%%%%%%%%%%%%
%
\item[-]
$J^\eta=|F|^2$ becomes infinite at $\xi=\xi_c$.  The current looks like a 
``jet stream''. It is due to the deflection of the incoming flux by the field 
during infinite time.
\end{itemize}

\section{Discussion}

After obtaining the positive result with the positronium model, the absence 
of transverse polarization in the Schwinger mechanism was rather unexpected. 
This absence does not happen due to standard discrete symmetries like C, P 
and T but due to invariance with respect to the particular transformation 
(\ref{m-q}). 

In spite of the difference between the positronium dissociation and the 
Schwinger mechanism, both models predict that the electron and the positron 
have equal spin components $s^+_x$ and $s^-_x$ along the $\hat x$ axis~:
\beq
s^+_x s^-_x = + 1/4
\label{correl}
\eeq
This property is built-in in the \3p0 positronium model.  For the Schwinger 
mechanism, both $s^+_x$ and $s^-_x$ are equal to ${1\over2}\beta\Sigma_x$. 
Indeed, a positron at rest has $s^+_x=-{1\over2}\Sigma_x$, since it is a 
``hole'', and $\beta=-1$ because the corresponding (unoccupied) state has 
negative energy.  Eq.~(\ref{correl}) imply that the pair is in a triplet 
state also for the Schwinger mechanism.

The difference between the two models may be connected with their different 
chiral properties, which appear in the $m\to 0$ limit~:
\3p0 positronium dissociation is similar to the decay of a $0^{++}$ particle 
into two fermions, in which the fermions necessarily have opposite chirality 
(equal helicity). On the contrary, the Schwinger mechanism involves only vector 
interactions, therefore $e^+$ and $e^-$ must have the same chirality (opposite 
helicity).  The difference is particularly important at zero transverse 
momentum, where conservation of angular momentum $J_z$ requires 
$s_z^- = - s_z^+$.  Positronium dissociation is allowed for $m=q_\perp=0$,
whereas pair creation (with back-to-back $e^+$ and $e^-$) is 
forbidden\begin{footnote}
%------------------------------------------------------------------
{Eq.(\ref{expo}) seems to allow pair creation at $m=q_\perp=0$. 
However, this is a too special case where
the field (\ref{champ}) occupies an infinite domain of the $(z,t)$ plane. 
In fact, for $m=q_\perp=0$, the functions $F$ and $G$ decouple 
(see Eqs. \ref{F},\ref{G}) and the left- and right-moving currents 
$J^\xi = |G|^2$ and $J^\eta = |F|^2$ are separately conserved.
If the field domain is finite in the $(z,t)$ plane,
$e^+ e^-$ pairs are produced at $m=q_\perp=0$,
but with $e^+$ and $e^-$ going in the {\it same} direction.
$e^+$ and $e^-$ going in opposite direction belong to independant pairs.
}\end{footnote}.
%-------------------------------------------------------------------

The absence of correlation between spin and transverse momentum in the 
Schwinger mechanism does not preclude such correlations for $q\bar q$ pairs 
created during string breaking. There are many effects of string breaking not 
included in the Schwinger mechanism. First of all, the chromoelectric field 
between the quark and the antiquark is totally screened after their creation, 
unlike in the Schwinger process where the field extends everywhere all the 
time\begin{footnote}
%---------------------------------------------------------------
{Let us mention however Ref.\cite{Kluger}
where the screening effect has been considered for scalar QED.} 
\end{footnote}.
%----------------------------------------------------------------
Secondly, the field of the QCD string is confined to a thin tube. One way to 
simulate this fact in QED is to impose the MIT-bag boundary conditions for 
the electron in the transverse coordinates \cite{Schonfeld,Pavel}.  In our 
problem, we cannot use this method because we need a well-defined transverse 
momentum.  Important effects may also come from the "transverse inertia" of 
the string, because part of it must follow the transverse motion of the quark.
Finally, the Schwinger mechanism does not include the final state interactions
which recombine the quarks from different pairs to form hadrons and resonances. 
Resonances probably play a major role in single spin asymmetries 
\cite{Barni,Collins94,Jaffe}, because the latter come from the interference 
between different spin amplitudes having different phases.

To summarize, the interplay of spin and transverse momentum in pair creation 
is a subtle phenomenon and we cannot conclude from the simple model presented 
above whether the correlation exists or not.

\section*{Acknowledgements}
We acknowledge the financial support from the IN2P3--Poland scientific exchange 
programme (collaboration no.~91-62). J.C. has been also supported by the Polish 
State Committee for Scientific Research (KBN) grant no.~2~P03B~086~14 and by 
the Polish-German Collaboration Foundation grant FWPN no.~1441/LN/94 during 
complition of this work.

\end{document}